# Permafrost – An Alternative Target Material for Ultra High Energy Neutrino Detection?

R. Nahnhauer[1], A. A. Rostovtsev[2], D. Tosi[1]

[1] Deutsches Elektronen-Synchrotron, DESY,
    Platanenallee 6, D-15738 Zeuthen, Germany
[2] Institute for Theoretical and Experimental Physics, ITEP,
    B. Cheremushkinskaya 25, 117218 Moscow, Russia

**Abstract:**

The detection of cosmic neutrinos with energies above $10^{17}$ eV got growing interest during recent years. Possible target materials for in-matter arrays of ~100 km$^3$ size under discussion are water, ice and rock salt. Here we propose to investigate permafrost as an additional alternative, covering ~20% of Earth land surface and reaching down to more than 1000 m depth at certain locations. If sufficiently large attenuation lengths for radio and acoustic signals can be demonstrated by in-situ measurements, the construction of a large hybrid array within this material may be possible in the Northern hemisphere. Properties and problems of a possible location in Siberia are discussed below. Some acoustic data are compared to laboratory measurements using "artificial" permafrost.

**Motivation:**

During recent years the detection of cosmic neutrinos of ultra-high energies got growing interest /1,2,3/. In particular the interaction of highest-energy charged cosmic particles with the cosmic microwave background should be a guaranteed source of neutrinos with energies above $10^{17}$ eV  (GZK neutrinos). Flux predictions are small and uncertain by about an order of magnitude /4/. It is clear, however, that large detection volumes have to be scanned to observe a handful of events. Present attempts to look either at the Moon surface with Earth based radio telescopes /5/ or to large ice volumes on Earth from satellites /6/ or balloons /7/ have not reported positive results until now. The problem remains to clearly separate the few expected signal events from the  background signals.

To diminish such problems, large volume in-matter hybrid arrays have been proposed, using at least two techniques with different systematics for background reduction and signal identification /2,3/.  For water, acoustic detectors are suggested to join and enlarge the optical KM3Net array /8/. First in-situ studies will be made in the Mediterranean within the ANTARES experiment /9/. Ice as target material has been proven to work excellently for optical and radio detection methods within the AMANDA, IceCube /10/ and RICE /11/  experiments. The acoustic properties of the Antarctic ice are being studied within the SPATS effort /12,13/. A 100 km$^3$ scale detector simulation using all three techniques at the South Pole promises the detection of about 10 GZK neutrino events per year with radio and acoustic sensors at the same time/14/. Finally, rock salt allows good radio detection of neutrino signals which is used by the SalSA collaboration to evaluate a large salt dome array /15/. This could be joined by acoustic detectors used already by geophysicists to study salt mine properties /13,16/.

All three materials and corresponding projects have, however, to handle a number of problems. For a water target in the open sea the biggest handicap is the large and unstable background from natural and human sources. The radio detection method does not work for water and optical and acoustic methods have very different energy thresholds. This makes it hard to construct a real hybrid array with only these two techniques. A good option seems to be an in-ice array with all three detection methods working in parallel. The present location in Antarctica however makes the construction expensive and time consuming. A corresponding second array in the northern hemisphere, necessary to cover the complete sky could only be located in Greenland. Until now nobody has studied such an option seriously, but the Greenland ice properties seem to be less favourable due to its higher temperature. The R&D program for a salt detector is strongly harmed by the huge start-up costs for drilling down to reasonable depth.

A further alternative target material available in the northern hemisphere is therefore an interesting option to be developed. A candidate for a such new material is permafrost since:

- The material is available at several locations in Canada, Russia and the US (Alaska) with depths going from several hundred meters down to thousand meters with reasonable extensions.
- Drilling in a sand-ice-water mixture should not be too complicated and is at least known to many oil- and gas-producing companies.
- Both radio and acoustic detection methods work in permafrost with about the same energy threshold allowing real hybrid event detection.
- The high density of the material will provide about twice the neutrino interaction rates of water and ice.
- The temperature below 0° C allows to bring acoustic detectors easily in contact with the target material just filling open holes with water after deployment.

**Permafrost properties:**

There is no clear definition of permafrost. Often it is considered to be a sub-surface layer with temperatures steadily below 0° Celsius, independently of the real material and structure. Typically it is represented by frozen alluvial soil or sediments resting on top of a rock platform. For a neutrino detector only the upper alluvial layer is of interest here. A further difficulty is the large inhomogeneity of the material near to the surface, due to temperature changes and climate. This region has definitely to be excluded from all further considerations. What remains is hopefully a mostly homogeneous material of fine grain sandstone from 50-100 m below surface to a depth as large as possible. Its properties are still dependent on grain size, ice-water content, mineral admixtures and probably several other parameters.

In this spirit a specific location was studied. With the help of the experts of the Permafrost Institute of the Siberian branch of the Russian Academy of Science in Yakutsk, the Aldan River valley, near to the entry of this river to the Lena River and not too far away from Yakutsk, was found to be a promising location (see fig. 1) /17/. No core from a bore hole in that region or other corresponding information was immediately available. The profile of a historical shaft in Yakutsk reaching to a depth of 116.5 m is shown instead in fig. 2 /18/. As can be seen, sandstone starts to be dominant at a depth of about 30 m being interrupted by layers of slate between 60 and 80 m only. The temperature in that shaft has been measured to be constant over more than hundred years with about -6° C near the surface and -3° C at the bottom.

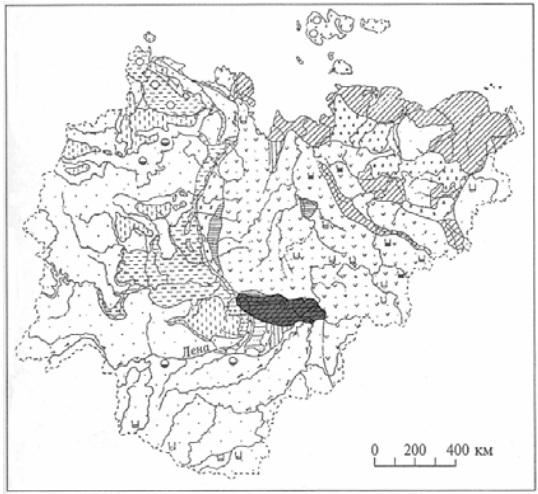

Fig. 1: Geological map of Yakutia with details about different permafrost regions. the dark cross hatched region should be dominantly sandstone with a depth between 150 m and 1060 m. (from /17/)

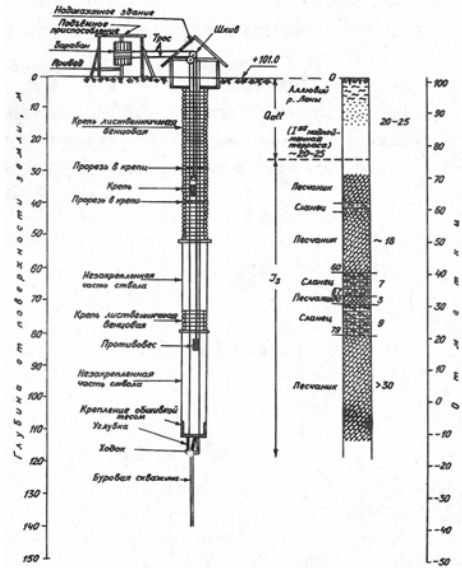

Fig. 2: Schergin well, a historical well in Yakutsk, excavated to a depth of 116.5 m in 1837 (from /18/ )

The content of non-frozen water in permafrost varies with temperature and soil dispersion. Whereas in permafrost sand 100% of the water is frozen at -10° C, permafrost clays contain a substantial fraction of non-frozen water even at -20° C. All this will influence the velocity of sound in the material . In /19/ this has been studied for different sand-water-ice mixtures, fig. 3 shows an example result. For a 20% ice content the velocity of sound comes out to be larger than 4000 m/s at temperatures below -3° C and is still larger than 3000 m/s for only 10% of ice admixture. Such values promise large pressure signals from neutrino interactions because in the thermo-acoustic model the pressure depends on the sound velocity squared. Other elastic properties of the material have still to be evaluated.

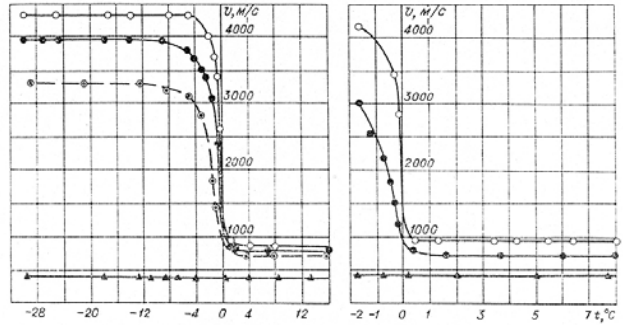

Fig 3: Velocity of ultra-sound in permafrost in dependence of the temperature for different water admixtures – open circles: 21%, closed circles: 16%, pointed circles: 10%, triangles 1% (from /19/).

**Laboratory measurements:**

For radio detection proof-of-principle measurements have been done for silica sand /20/ and ice /21/, so one may fairly assume, that a corresponding mixture will also give positive results. Acoustic accelerator studies have extensively been done for water /22/ and ice /23/, but no results for sand are available until now. That was the reason for an experimental check of acoustic properties of artificial "permafrost" in the laboratory. For that purpose a mixture of ~ 80% sand and 20% water was filled in the setup drawn in fig. 4, mixed and frozen.

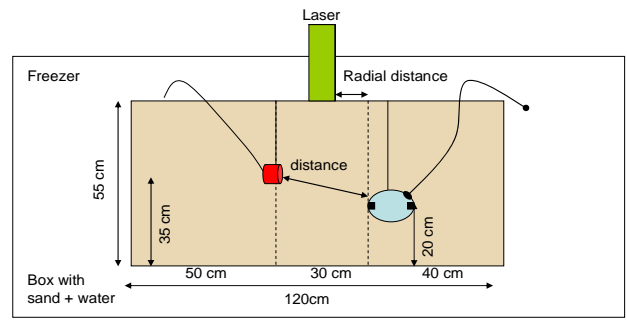

Fig. 4: Setup to measure acoustic properties of artificial permafrost. Red: piezo transmitter, blue: two channel acoustic sensor, green: YAG laser.

After freezing a density of 2.0 ± 0.1 g/cm$^3$ was measured for the material. A piezo transmitter was used to send sinusoidal pressure waves of different frequency to a pair of acoustic sensors over a longitudinal distance of 30 (40) cm. A YAG laser with a 1mm wide beam of an energy of 10 mJ/pulse and a pulse length of 6ns could be moved along the setup to create a thermo-acoustic signal at variable distances. The temperature of the target could be changed between -5° C and -30° C.

In a first measurement the velocity of sound was measured using the piezo transmitter at a frequency of 25 kHz and an input voltage of 2V. As can be seen from fig. 5, in the considered temperature range a speed of sound of $v_s = 4000 \pm 200$ m/s was found in good agreement with the permafrost result shown in fig. 3.

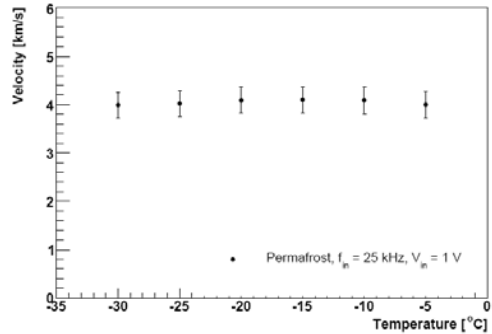

Fig. 5: Velocity of sound in dependence of temperature for 25 kHz sound waves in artificial permafrost.

In fig. 6 the dependence of the signal amplitude on the transmitter input voltage at 25 kHz is shown for "permafrost" at -20° C in comparison to water at 18.5° C. For both materials a linear dependence is observed but the signal measured in permafrost is ten times larger than in water. As can be seen in fig. 7, the signal is nearly independent of the permafrost temperature as also expected from results for the natural material.

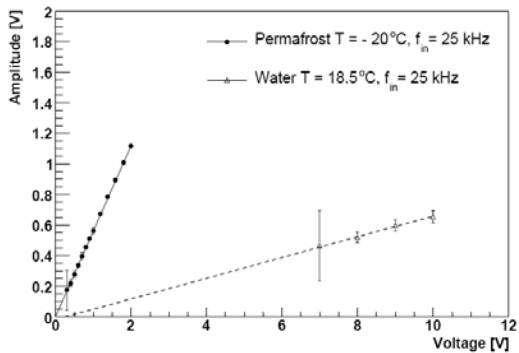 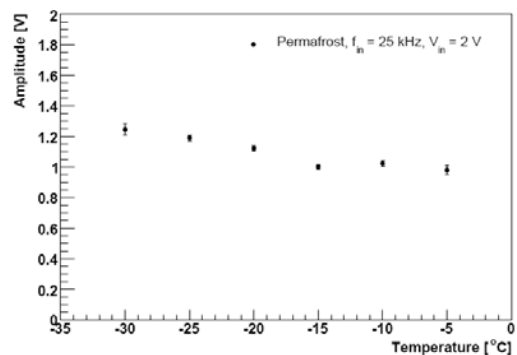

Fig 6: Signal amplitude at receiver in dependence of piezo-transmitter input voltage for artificial permafrost and water.

Fig 7: Temperature dependence of sound wave signal amplitude in artificial permafrost.

The dependence of the signal amplitude on the input frequency is shown for permafrost and water in fig. 8. The observable peak frequency depends in both cases on the detector construction. For "permafrost" the signal reaches its maximum at about 50 kHz whereas for water a better sensitivity is measured at higher frequencies. Because the thermo-acoustic neutrino signals are expected to be produced dominantly in the 20 to 50 kHz range  the presently used sensors are favourable for their detection in permafrost.

To measure the variation of the signal with distance d, the laser was moved in steps of 5 cm  on top of the target material. In principle this should produce a thermo-acoustical pressure pulse giving rise to a cylindrical wave. However the absorption length of the laser light in the used material is not known but probably very short, which finally may cause only a point like excitation near the surface. The result of the measurement is shown in fig. 9. Above 15 cm  the data are compatible with a 1/d behaviour. Further conclusions are not possible to be drawn due to measurement errors and the small available distance range.

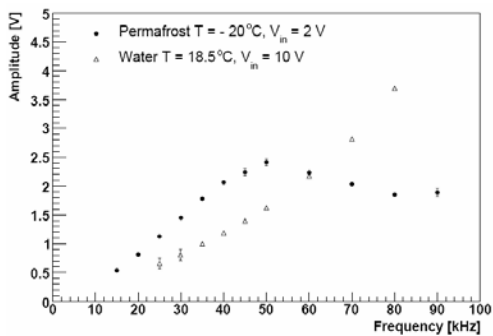

Fig 8 Dependence of signal sound wave amplitudes  on the  input frequency for artificial permafrost and water

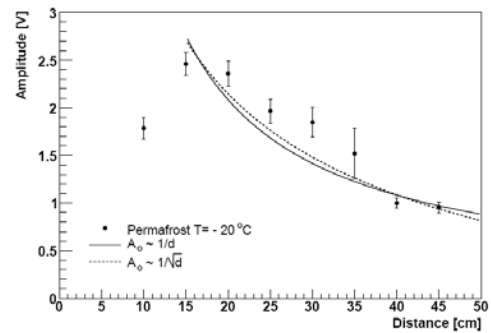

Fig 9 Dependence of sound wave signal amplitude on the distance from the source for artificial permafrost.

**Conclusions and outlook:**

From its mechanical properties the frozen sand-ice-water mixture called permafrost throughout this paper seems to be appropriate as target material for a large radio-acoustic hybrid neutrino detector. The main open question to be solved is to determine the attenuation lengths for radio and acoustic signals in the 100-1000 MHz and 10-100 kHz frequency range, respectively.  These parameters will strongly depend on the geological structure at the chosen location. A region in Siberia near Yakutsk has been identified to be promising, being large enough to house a future real detector of reasonable size, if necessary in-situ measurements would give positive results. To do such investigations, requires the common effort of a group of geophysicists and astro-particle physicists within a TAIGA[*] exploration study .

---

[*] Test Askaryan Installation for Geology and Astrophysics

**Acknowledgement:**

The authors are deeply indebted to V. N. Efremov, G. P. Kuzmin and V. V. Shepelev from the Yakutsk Institute of Permafrost of the Russian Academy of Science (Siberian Branch) for teaching us several permafrost details as well as providing information about permafrost conditions in Yakutia. Furthermore we thank our colleagues from the Yakutsk EAS array experiment in particular M. I. Pravdin and V. I. Kozlov, for introducing us to their experiment and parts of the Yakutian infrastructure.

**References:**

/ 1/ "Radio detection of high energy particles",
    Proceedings of the "First International Workshop RADHEP 2000,
    Los Angeles 2000, Editors: D. Saltzberg and P. Gorham

/ 2/ "Acoustic and Radio EeV Neutrino Detection Activities"
    Proceedings of the workshop "ARENA 2005"
    Zeuthen, 2005, Editors: R. Nahnhauer and S. Böser

/ 3/ "Acoustic and Radio EeV Neutrino Detection Activities"
    Proceedings of the conference "ARENA 2006"
    Newcastle 2006, to be published

/ 4/ R.Engel, D. Seckel and T. Stanev, Phys.Rev. D64 (2001) 093010

/ 5/ P.W. Gorham et al. Phys Rev. Lett. 93, 041101 (2004)

/ 6/ N. G. Lethinen et al., Phys. Rev. D69, 013008 (2004)

/ 7/ "Detection of Ultra High Energy Neutrinos via Coherent Radio Emission"
    G. S. Varner et al., SLAC-PUB -11872, presented at SNIC 2006, Menlo Park, 2006

/ 8/ "Detection of UHE neutrinos with an underwater very large volume array of acoustic sensors/ A simulation study"
    T. Karg, astro-ph/0608312 (2006)

/ 9/ "Integration of Acoustic Neutrino Detection into ANTARES",
    K. Graf et al, astro-ph/0703442, to appear in / 2/

/10/ E. Andres et al., Astropart. Phys. 13 1 (2000)
    A. Achterberg et al., Astropart. Phys. 26 155 (2006)

/11/ I. Kravchenko, et al. Astropart. Phys. 20, 195 (2003)

/12/ "Feasibility of acoustic neutrino detection in ice: Design and performance of the South Pole Acoustic Test Setup (SPATS)"
S. Böser et al., to appear in Proceedings of the ICRC-2007, Merida, Mexico (2007)
"Feasibility of acoustic neutrino detection in ice: First results from the South Pole Acoustic Test Setup (SPATS)"
S. Böser et al., to appear in Proceedings of the ICRC-2007, Merida, Mexico (2007)

/13/ P. B. Price, J. Geophys. Research, V 111 (B02201) 2006

/14/ "Simulation of a Hybrid Optical/Radio/Acoustic Extension to IceCube for EeV Neutrino Detection"
D. Besson et al., Proceedings of the ICRC-2005, V 5, 29, Pune 2005 and astro-ph/0512604

/15/ "Introduction to SalSA, a saltdome shower array as a GZK neutrino observatory"
D. Saltzberg et al., in / 1/ page 252

/16/ "Experience on acoustic wave propagation in rock salt in the frequency range 1-100kHz and conclusions to the feasibility of a rock salt dome as neutrino detector"
J. Eisenblätter et al., in / 1/ page 30

/17/ G.P. Kuzmin, Poolemnye sooruzhenia v kriolitzone
Novosibirsk, Nauka, 2002

/18/ P.A. Solovjov, Schachta Schergina, AN SSSR Sibirskaja Otdelenia, Yakutsk, 1982

/19/ I. N. Votjakov, Fisiko mechanicheskie svoistva merslych i ottoivajuschschich gruntov Jakutii, Novosibirsk, Nauka, 1975

/20/ D. Saltzberg et al., Phys. Rev. Lett.86, 2806 (2001)

/21/ "Observation of the Askaryan Effect in Ice"
P. W. Gorham et al., hep-ex/0611008, submitted to Phys. Rev. Lett.

/22/ G. Sulak et al. NIM 161(1979) 203
V.I. Albul et al., Instr. and Exp. Techn. 44 (2001)327

/23/ "Acoustic Sensor and Transmitter Development for a Large Volume Neutrino Detector Array in Ice"
S. Böser et al., Proceedings of the ICRC-2005, V 5, 29, Pune 2005